\documentclass{aa}
\usepackage{graphicx}
\usepackage{txfonts}
\usepackage{natbib}
\usepackage[english]{babel}
\bibpunct{(}{)}{;}{a}{}{,}
\newcommand{\HII}{ H\,{\sc{ii}}}

\begin{document}

\title{Spitzer/IRAC view of Sh\,2-284
\thanks{ Based on data obtained with IRAC onboard Spitzer (Program ID 3340) and the WFC at Isaac Newton Telescope at La Palma.}}
\subtitle{Searching for evidence of triggered star formation in an isolated region in the outer Milky Way}
\author{E. Puga\inst{1}
\and S. Hony\inst{2,1}
\and C. Neiner\inst{3,1}
\and A. Lenorzer\inst{4}
\and A.-M. Hubert\inst{3}
\and L.B.F.M. Waters\inst{5,1}
\and F. Cusano\inst{6}
\and V. Ripepi\inst{7}
}

\offprints{E. Puga}

\institute{Instituut voor Sterrenkunde, Katholieke Universiteit Leuven, Celestijnenlaan 200D, 3001 Leuven, Belgium\\
\email{elena@ster.kuleuven.be}
\and  Laboratoire AIM, CEA/DSM - CNRS - Universit\'e Paris Diderot, DAPNIA/SAp,
91191 Gif-sur-Yvette, France
\and  GEPI, Observatoire de Paris, CNRS, Universit\'e Paris Diderot; 5
place Jules Janssen, 92190 Meudon, France
\and Instituto Astrof{\'{i}}sico de Canarias,C/ V{\'{i}}a L{\'a}ctea s/n, E-38200, La Laguna, Spain  
\and Sterrenkunde Instituut, Universiteit van Amsterdam, Kruislaan 403, 1098SJ
Amsterdam, The Netherlands
\and Th\"uringer Landessternwarte Tautenburg, Sternwarte 5, D-07778 Tautenburg, Germany
\and INAF- Osservatorio Astronomico di Capodimonte, Via Moiariello 16, I-80131 Napoli, Italy
}
\date{Received ; accepted }

\abstract
{}
{Using Spitzer/IRAC observations of a region to be observed by the
  CoRoT satellite, we have unraveled a new complex star-forming region
  at low metallicity in the outer Galaxy. We perform a study of S284
  in order to outline the chain of events in this star-forming
  region.}
{We used four-band Spitzer/IRAC photometry as well as H$\alpha$
  imaging obtained with INT/WFC. Combining these data with the optical
  photometry obtained in the frame of CoRoTs preparation and the 2MASS
  catalog we analysed the properties and distribution of young stellar
  objects (YSOs) associated with point-like sources. We also studied
  the SEDs of regions of extended emission, complementing our dataset
  with IRAS and MSX data. }
{We find that S284 is unique in several ways: it is very isolated at
  the end of a spiral arm and both the diffuse dust and ionized
  emission are remarkably symmetric. We have partially resolved the
  central clusters of the three bubbles present in this region.
  Despite the different scales present in its multiple-bubble
  morphology, our study points to a very narrow spread of ages among
  the powering high-mass clusters. In contrast, the particular
  sawtooth structure of the extended emission at the rim of each
  ionized bubble harbours either small lower-mass clusters with a
  younger stellar population or individual young reddened protostars.
  In particular, triggered star formation is considered to be at work
  in these regions.}
]{}

\keywords{Stars: formation, pre-main sequence, ISM: bubbles, HII
  regions, ISM: individual: Sh 2-284, Galaxy: open clusters and
  associations: individual: Dolidze~25 }

\maketitle

\section{Introduction}
\begin{figure*}[!t]
  \includegraphics[width=\textwidth]{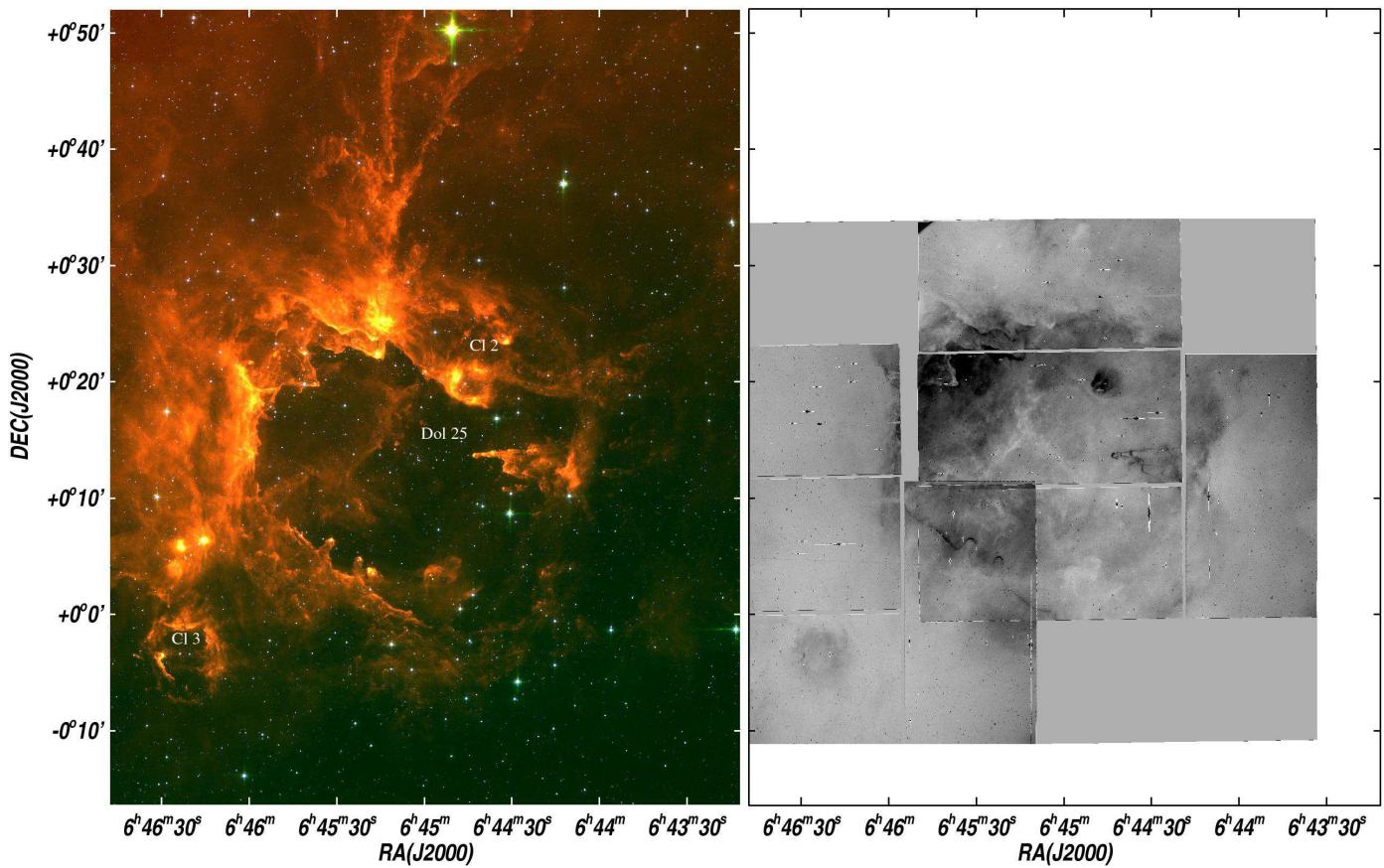}
  \caption{{\bf (Left panel)} Colour-composite image of S284 with
    IRAC ({\it blue}: IRAC\,3.6, {\it green}: IRAC\,5.8, {\it red}:
    IRAC\,8.0). {\bf (Right panel)} Inverted grey scale of the
    H$\alpha$ emission of S284 obtained with the WFC at the Isaac
    Newton Telescope. The image is in arbitrary units.}
  \label{S284_col_Ha_image}
\end{figure*}
Several scenarios have been put forward to explain the role of
expanding \HII{} regions in the frame of triggered star formation
\citep{Elmegreen,Deharveng}. Observationally, it is difficult to find
an isolated \HII{} region in the Galactic plane where the impact of
the expansion can be decoupled from severe crowding effects \citep[see
summary by ][]{Churchwell_06}. However, triggered SF is commonly
invoked when \HII{} regions and bubbles are aligned and a
multiple-bubble morphology is detected. Several examples of \HII{}
regions have been recently investigated a few degrees off the Galactic
disk plane as an illustration of the specific hypothesis called:
"collect and collapse mechanism"
\citep{Deharveng,Zavagno_06,Zavagno_07}. The outer region of our
Galaxy offers a benchmark to study this type of object
\citep{Ehlerova}. Away from the central spiral arms, the third
quadrant is sparsely populated with the giant molecular clouds
belonging to the Perseus and Norma-Cygnus arms. Moreover, lower
metallicities are measured toward the outer Galaxy
\citep{Martin-Hernandez}, guarantee of closer conditions to
the pristine initial molecular clouds.

Sh\,2-284 (hereafter S284) is a diffuse nebula observed for the
Sharpless (1959) catalog. This region is located in the anti-centre of
the Galactic disk (l=212\degr, b=$-$1.3\degr), and it is relatively
isolated. It shows a very symmetric morphology in the red plate of the
Palomar Observatory Sky Survey (POSS) with an approximated diameter of
$80\arcmin$ in the sky. The kinematical study of the H$\alpha$ line in
S284 revealed a two-component velocity profile \citep{Fich_90}. The
radial velocities of these components are 43.0$\pm$0.5 and
$-$20.1$\pm$0.5 km\,s$^{-1}$, while the linewidths are 48.6$\pm$0.5
and 26.4$\pm$0.8 km\,s$^{-1}$, respectively. Later studies of several
recombination lines at 9\,cm toward S284 \citep{Lockman} encountered
similar values for this first velocity component (V=45.2$\pm$1.5
km\,s$^{-1}$, $\Delta$V=28.2$\pm$3.6 km\,s$^{-1}$). The narrow
linewidths of the ionized gas tracers toward S284 are compatible with
a diffuse \HII{} region origin. These estimates of the radial
velocities agree with the values obtained from the observation of the
CO emission \citep[see ][]{Blitz,Wouterloot_89}. Radial velocities in
the range V$_{\rm{LSR}}$\,$\sim$42.0--46.8 km\,s$^{-1}$ have meen
measured at various locations in the field, yielding heliocentric
distances between 5.2 and 6.5 kpc.

\indent S284 hosts a cluster of stars in its centre known as
Cl~Dolidze~25 \citep{Dolidze}; aka C\,0642+0.03 or OCL-537. The
stellar population of this cluster was initially characterised through
UBV--H$\alpha$ photometry by \citet{Moffat_Vogt}. They concluded that
Dolidze~25 was a cluster of 10 O and B type stars and 1 F supergiant.
However, in their study of a 23\arcmin\,$\times$\,16\arcmin\,field,
the authors separated the cluster into two subclusters with different
reddening. A northern group of foreground stars of E($B-V$)$\approx$15
at a distance of 0.75 kpc and a redder and more concentrated central
cluster. \cite{Turbide_Moffat} have pinned down its reddening to
E($B-V$)$\approx$0.83 and the distance to 5.0 kpc, allowing for a
metallicity gradient in the Galactic rotation curve. These results
cast doubt on the physical connection between the different clusters
found in this field. Although much has been discussed about the actual
members of the cluster \citep[see][]{Babu,Turbide_Moffat,Pandey_06},
most authors agree on the distance to the central concentration of
stars. \cite{Russeil_07} recalculated this parameter using previous
UBV photometry and spectroscopy \citep{Moffat_79,Turbide_Moffat}
yielding a distance of 6.03$\pm$1.16 kpc. However, the selected stars
to derive this value do not belong to the central subcluster. A more
recent spectroscopic study of the sources Dol~25--15, Dol~25--17, and
Dol~25--22 within this central subcluster \citep{Lennon} the authors
determined: (i) earlier spectral types than prevously derived for
these sources, (ii) the low metallicity of the cluster
(Z$=$0.17\,Z$_{\sun}$), (iii) a spectro-photometric distance of
5.5$\pm$0.5 kpc. We adopt a distance of 5 kpc throughout the present
study. The age of the cluster (7 Myr) has been estimated through
isochrone fitting of precision photometry in the
4\arcmin\,$\times$\,4\arcmin\,field around Dolidze~25
\citep{Turbide_Moffat}.

Water-maser emission has been detected toward a few locations around
S284 \citep{Wouterloot_88,Brand_94}, whereas no SiO, CS, or methanol
maser emission (i.e. tracers of very dense ambient material) have been
detected \citep{Harju_98,Plume_92,Zinchenko_98,Bachiller_90}.

The objective of this paper is to determine the properties of the
different regions identified toward this site and their physical
interconnection. In particular, it focuses on scrutinizing the
evidence for sequential star formation.

This article is structured as follows. In Sect.~\ref{obs_data_red} we
describe the observations of this star-forming region (SFR), obtained
with Spitzer/IRAC (Sect.~\ref{IRAC}). We also present ground-based
H$\alpha$ observations obtained with INT/WFC (Sect.~\ref{INT}), as
well as additional optical data, which we cross-correlate with 2MASS
and Spitzer/IRAC. Secondly, we focus on the stars and distribution of
young stellar objects (Sect.~\ref{point_sources}), followed by an
analogous study of blobs of extended emission (Sect.~\ref{blobs}). In
Sect.~\ref{disc} we discuss the spectral types of the stellar content
in S284 (Sect.~\ref{spect+mass}) and the relative ages and timescales
we determine (Sect.~\ref{age+timesc}). In Sect.~\ref{environ} we
discuss the uniqueness of this nebula in terms of its particular
environmental conditions. Finally, conclusions are drawn in
Sect.~\ref{concl}.

\section{Observations and data reduction}\label{obs_data_red}
\subsection{Spitzer/IRAC Imaging } \label{IRAC}
\begin{figure}[!t]
  \includegraphics[width=8.8cm]{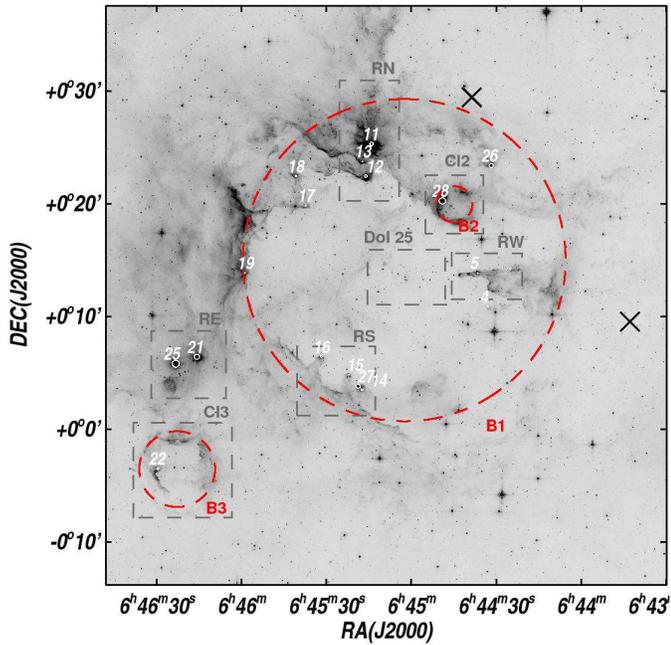}
  \caption{Map of S284 in the IRAC\,5.8 band in logarithmic scale.
    The bubbles are identified with large discontinuous circles to
    guide the eye, blobs are represented by small circles, and
    discontinuous boxes indicate the selected regions for close-up
    study. The x-symbols correspond to the H$_2$O maser emission
    reported by \cite{Brand_94} and \cite{Wouterloot_88}. }
  \label{box_blobs}
\end{figure}
We observed S284 with the IRAC (InfraRed Array Camera) onboard Spitzer
during Cycle 4 (2005 March 28) under Programme ID 3340. The
observations were conducted in the four IRAC filters at 3.6, 4.5, 5.8,
and 8.0 $\mu$m \citep{Fazio}. S284 spreads over an area of
1\degr\,$\times$\,1\fdg4 with its centre at
$\alpha_{2000}$$=$\,06$^h$\,45$^m$\,00\fs0,\,
$\delta_{2000}$$=$\,$+$00\degr\,17\arcmin\,56\farcs9\,. We performed a
raster map with IRAC mapping mode in each filter with a spacing
between rows of $1/3^{\rm rd}$ of the camera field of view
(5\farcm2\,$\times$\,5\farcm2). The IRAC simultaneous imaging in two
consecutive filters results in an offset for two of the mosaic fields
of around 6\farcm8 (in our case, in declination). A final area,
covered in all filters, of approximately 0\fdg95\,$\times$\,1\fdg3 was
obtained. We obtained an average integration time of 80.4 seconds per
sky pixel.

The Spitzer Science Center provided the Spitzer/IRAC basic calibrated
data (BCD), using the pipeline version S11.4.0. Further treatment of
the images with the contributed IRAC software was necessary to correct
for the muxbleed and pulldown effects of the detector. Likewise, the
difference in the background levels of overlapping frames present in
the IRAC 8.0\,$\mu$m channel was corrected with the post-BCD overlap
correction software. Subsequently, we used the dedicated software
package MOPEX (version 030106) to produce mosaics of the IRAC images
and to perform PSF photometry \citep{Makovoz}.

MOPEX performs image interpolation and co-addition of the individual
images onto a common grid. Moreover, it allows outlier rejection
through four different algorithms. Considering the shallow coverage of
the field, cosmic rays and other outliers were removed using the box
outlier and dual outlier MOPEX modules. The mosaic final pixel scale
corresponds to the original 1\farcs22\,/pix. An image of this field in
three IRAC colours is shown in Fig.~\ref{S284_col_Ha_image}.

The source extraction is preceded by the construction of our own point
response function (PRF i.e., an oversampled PSF). We selected 30
isolated bright stars by hand from the final mosaic for each filter.
The postage stamp images of these sources were resampled to 1/4th. We
normalized the PRF for a radius that encompasses the first Airy ring
for each band, therefore an aperture correction needed to be applied
for the extracted flux on each single channel. The presence of strong
filamentary extended emission in the four bands required the use of a
small-width median filter for a proper background subtraction
(11\arcsec for IRAC\,3.6 $\mu$m and IRAC\,4.5 $\mu$m and 6\farcs35 for
IRAC\,5.8 $\mu$m and IRAC\,8.0 $\mu$m). This aggressive filtering was
accordingly applied to the channel PRF, producing an underestimate of
the source flux. The correction factor was determined from the flux
ratio between the unfiltered and filtered PRFs \citep{Shenoy}.

The source detection was performed in the single final coadded frame
for each individual channel. To ensure a high reliability, we only
selected sources with S/N\,$>$\,5. We finally obtained extraction
lists of 124\,657, 118\,289, 19\,445 and 15\,856 sources in the
IRAC\,3.6, 4.5, 5.8, and 8.0 $\mu$m channels, respectively (hereafter
referred to as IRAC\,3.6, IRAC\,4.5, IRAC\,5.8, and IRAC\,8.0). The
final merged list of the four bands was obtained by matching the
coordinates of the extracted sources in the four bands (using an
optimally determined maximum distance of 1\farcs5). To minimize the
contamination by spurious detections in the source catalog we apply a
filtering criterion based on simultaneous detection in multiple
filters. We require detection of the object in either the two bluest
(IRAC\,3.6 and 4.5) or reddest bands (IRAC\,5.8 and 8.0). This
criterion accomodates faint blue sources and very red sources allowing
them to survive into the source list. A total of 8\,010 sources were
extracted in the four filters. Since the IRAC/Spitzer observations
were aimed at relatively faint sources, very bright objects appear
saturated in our images. We determine the saturation limit of each
band by comparison of the fluxes obtained with PRF fitting and
aperture photometry (aperture radius of 3 pixels). The limit is set by
eye at the beginning of the nonlinear regime between these two fluxes.

\subsection{Cross-correlation with 2MASS }
In addition to the Spitzer/IRAC we used the 2MASS catalog. We
cross-correlated the point sources found in the IRAC images with this
catalog to obtain $J$-, $H$-, and $K_s$-band photometry. At least one
of these colours was found in 2MASS for 23\,734 point sources at a
distance closer than 1\farcs5.

\subsection{Cross-correlation with INT/WFC}\label{INT}
Since the field containing S284 is a target field of the CoRoT
satellite, it has been studied by CoRoT's exoplanet team led by M.
Deleuil \citep{deleuil}. As a preparation for the CoRoT mission, this
team obtained 6187 images in CoRoT's fields with the WFC (Wide Field
Camera) at the 2.5m INT (Isaac Newton telescope) in La Palma (Spain),
in order to classify the stars and select the best candidates for
planetary transits. They used the $B$ and $V$ Bessel filters and the
{\it r'} and {\it i'} Sloan Gunn filters. See Deeg et al., in
preparation for more details. The exoplanet team of CoRoT kindly put
at our disposal the photometry they derived in these four filters from
their observations for the field of S284. The mean seeing of the data
for this field is about 1.5\arcsec and the mean astrometric error is
0.06\arcsec.

In addition, we have obtained specific observations of S284 in 2005
December 18 with the same instrumentation (WFC/INT) with a focus on
the extended emission. We used the H$\alpha$ and ($\lambda$ 6568)
Sloan Gunn {\it r'} filters to trace the recombination line and
continuum, respectively. Two pointing positions were needed to cover
the nearly 1\degr\,$\times$\,1\degr\,region. We obtained two exposures
of 1200 s in H$\alpha$ and 300 s in {\it r'}. After bias, flatfield,
and astrometry corrections, the 4 different CCDs were independently
corrected for optical distortions with a $3^{\rm rd}$ order
polynomial. The different exposures and pointings were combined with
the IRAF routine MSCRED. Subsequently, the stellar continuum was
subtracteed in order to obtain a line emission image. After exposure
time synchronization, the {\it r'}-band mosaic was multiplied by the
flux ratio of 23 stars in the field in both bands. Since we are
interested in the extended emission, no absolute flux calibration was
performed on the data. The H$\alpha$ (continuum subtracted) image is
shown in the right panel of Fig.~\ref{S284_col_Ha_image}.

\section{Results}\label{res}
\begin{figure}[!t]
  \includegraphics[width=8.8cm]{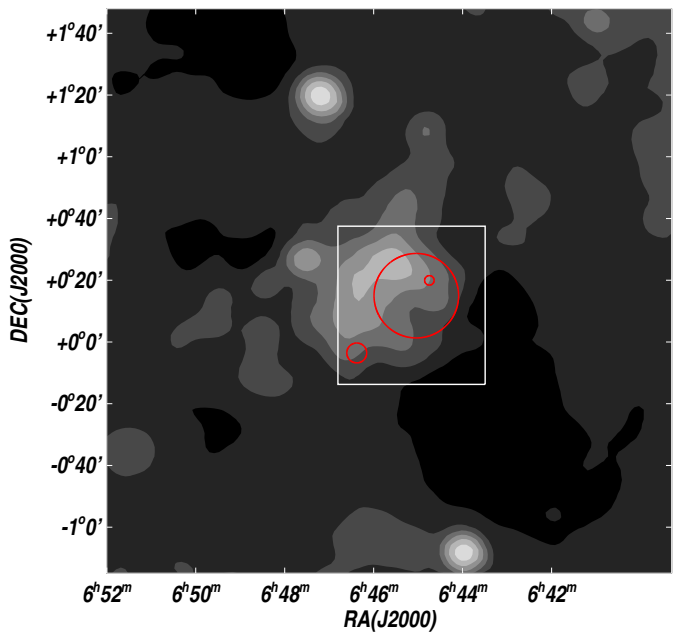}
  \caption{ Map of the A$_{V}$ extinction in the
    3\degr\,$\times$\,3\degr\,region around S284. The logarigthmic
    filled contours correspond to A$_{V}$ $=$[1.5, 2.25, 3.3, 4.7,
    6.9, 10.0, 14.5, 21.1] mag. The white box corresponds to the field
    of view shown in Fig.~\ref{box_blobs}. The bubbles identified in
    S284 are represented with circles. }
  \label{dust_extinction}
\end{figure}
\begin{figure}[!t]
  \includegraphics[width=8.8cm]{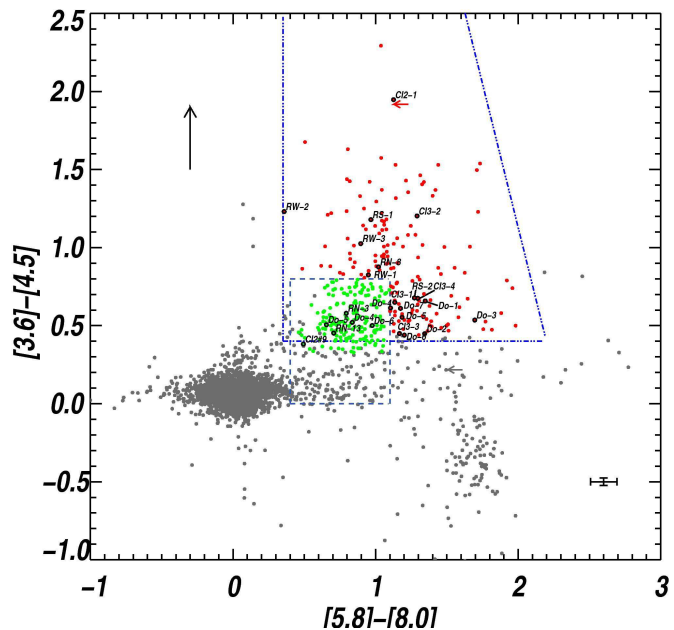}
  \caption{IRAC colour-colour diagram of S284. The dots represent the
    objects that are detected in the four IRAC bands, while arrows
    correspond to the limiting position of objects that appear
    saturated in the IRAC\,8.0 band. Median photometric errors are
    indicated with the error bars in the bottom right corner. The
    boundaries of the regions for Class~I and Class~II objects are
    taken from \cite{Allen04}. The black arrow represents a reddening
    vector for a visual extinction of 30 mag.}
\label{CI-CII_total}
\end{figure}
\begin{figure*}[!t]
  \includegraphics[width=8.8cm]{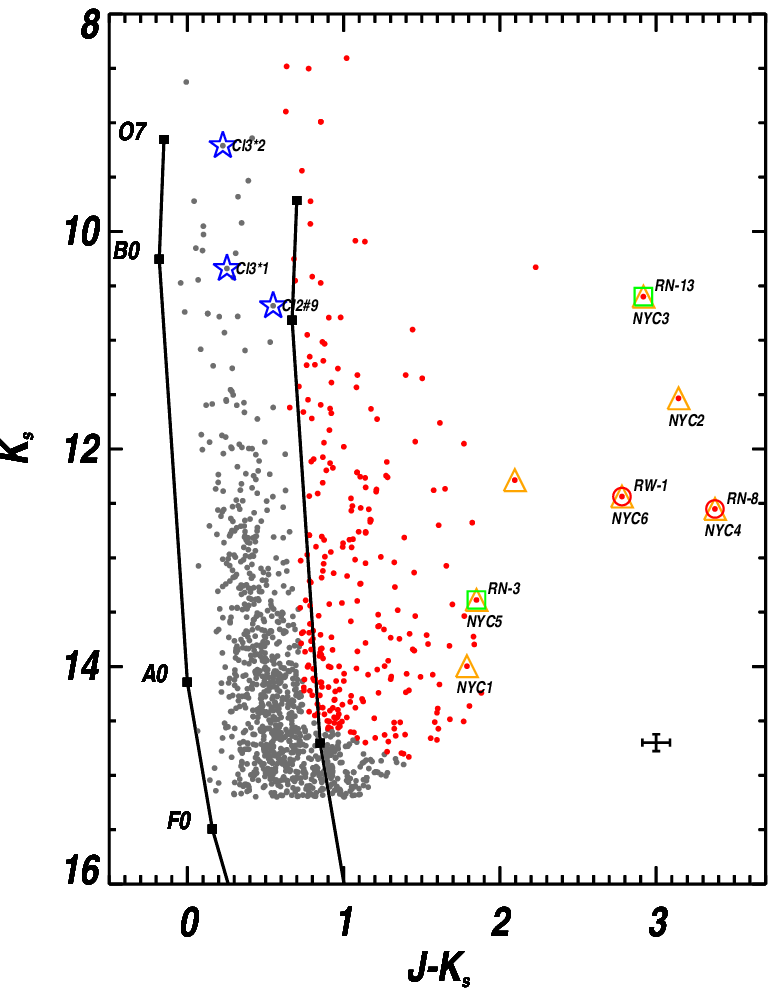}
  \includegraphics[width=8.8cm]{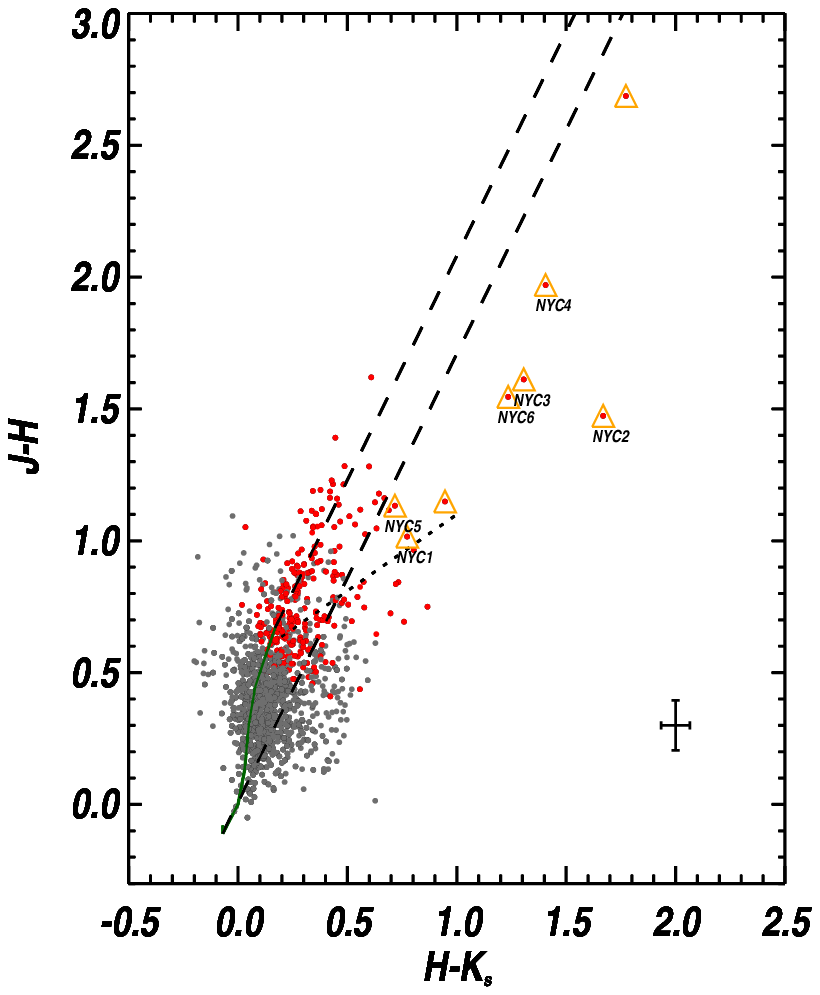}
    \caption{{\bf (Left panel)} 2MASS colour-magnitude diagram of
      the stellar content within the square areas indicated in
      Fig.~\ref{S284_col_Ha_image}. The lines connecting the small
      squares represent the main sequence of luminosity class V stars
      without extinction and with A$_{V}$ $\sim$ 5 mag. The labels
      correspond to sources further discussed. {\bf (Right panel)}
      2MASS colour-colour diagram of the same regions. The
      dashed lines indicate the direction of extinction according to
      an $\alpha$ $=$1.61 extinction law \citep{Rieke&Lebofsky}. The
      dotted line represents the T~Tauri locus of \cite{Meyer}.
      Therefore, the triangles highlight sources with near-IR excess
      (YSO candidates). Median photometric errors are represented by
      crosses in the lower-right corners of both diagrams [see the
      electronic edition of the Journal for a colour version of this
      figure].}
    \label{NIR_CMD_CCD}
\end{figure*}
Large dust bubbles are ubiquitous in the Galactic plane toward the
Galactic centre as shown by the GLIMPSE Survey \citep{Churchwell_06}.
However, the study of these structures is being hampered by the
difficulties imposed by crowding effects, the strong background toward
the Galactic midplane, and the frequent asymmetries in the bubbles as
a result of interactions with closeby regions.

The unprecedented view of S284 in the mid-IR provided by IRAC/Spitzer
(see Fig.~\ref{S284_col_Ha_image} {\it left panel}) has revealed in
detail the presence of complex extended emission in this region, which
was not previously detected in 2MASS observations. The emission is
present in all four IRAC bands implying diffuse dust emission. This
interpretation is corroborated by its detection (although at much
poorer resolution) in IRAS maps, even at 100 $\mu$m. The mid-IR
emission appears the strongest in the IRAC\,8.0 band, since this band
contains the PAH band emission at 7.7 $\mu$m and 8.6 $\mu$m. On the
largest scale toward S284, the images reveal that the dust shells
bound the bright H$\alpha$ emission quite well (see also
Fig.~\ref{S284_col_Ha_image} {\it right panel}), therefore tracing the
photon-dominated regions (PDRs) and the dense swept-up shells. These
two complementary maps allow us to identify three bubble-like
structures shown in Fig.~\ref{box_blobs}. In fact, the IR emission
extends well beyond the 18-pc inner bubble radius of the largest
bubble (hereafter B1). The dust emission almost fully surrounds the
ionized gas although it appears less prominent to the southwest,
likely due to a density gradient within the molecular cloud. This is
also evident on the much larger scale traced by the \cite{Schlegel}
Galactic dust extinction map (see Fig.~\ref{dust_extinction}), where
lower extinction values are detected in this direction. In the same
direction, the H$\alpha$ map shows a faint extended emission
component. This leak of ionizing photons suggests that the \HII{}
region may be density-bound to the southwest. The OB cluster
Dolidze~25, as outlined by \cite{Turbide_Moffat} and \cite{Babu}, is
located at the geometrical centre of B1. Interestingly, our H$\alpha$
image (Fig.~\ref{S284_col_Ha_image} {\it right panel}) traces the
bright rim of dense ionized gas, especially bright to the
southeast of the ionized bubble.

The resolved filamentary structure of the dust allows differentiating
among several other regions in the field of S284. Another two shells
of dust enclose more compact \HII{} regions within the field around
S284. The bubble B2 is located to the northwest, sitting amidst the
ionized emission of B1 and yet very close to its rim. This compact
bubble (diameter$\sim$ 3.5 pc), seems associated with IRAS06422+0053
which is the brightest point-source at 100\,$\mu$m. The low extinction
observed toward this area of ionized gas, together with the higher
brightness in H$\alpha$ indicate that it is probably located in the
foreground of B1. We detect a stellar cluster in its centre labelled
as Cl2 (see the left panel in Fig.~\ref{S284_col_Ha_image}). Two of
the sources of Cl2 belong to the trail of stars that has often been
associated with Dolidze~25.

A third bubble of 4-pc diameter lies within the mid-IR emission and is
located to the southeast of B1. The ionized gas is barely detected in
the INT H$\alpha$ mosaic. This source coincides with IRAS06439-0000
and IRAS06439+0111. We report the detection of another stellar cluster
(labelled as Cl3) at the centre of B3. Overall, these three bubbles
show a remarkable symmetry despite the highly probable interaction
among them.

Another prominent type of structure observed toward S284 are several
cometary globules that are detected at the rim of bubble B1. These
globules appear in emission in the IRAC images and as absorption
silhouettes in the H$\alpha$ image. Moreover, they appear highlighted
in the latter map by the presence of ionization rims at the head of
the globules. All of them point toward the centre of B1 showing very
diverse stellar content (see next section) and morphology with their
location around the nebula. Globules to the northeast show wide tails
and harbour resolved clusters of sources. They also seem to be located
at various radial distances from Dolidze~25, possibly due to a
projection effect (e.g. the region labelled as RN in
Fig.~\ref{box_blobs}). The cometary globules or elephant trunks to the
south and west possess longer and thinner tails and only contain a few
unresolved sources at the head, in IRAC\,3.6 and 4.5.

As measured by the shielded material behind the condensations, this
distinctive morphology between cometary globules and elephant trunks
supports our hypothesis of a density gradient in the ambient material.
However, observations
of the molecular gas are needed to confirm this statement.

Bubble B3 shows only one cometary globule at the eastern rim of the
dust bubble. A resolved source appears in the head of this cometary
structure, which points to cluster Cl3. Some of these globules and
trunks are associated with isolated IRAS sources, indicating an
enshrouded nature. Discussion on their stellar content follows in the
next section.

\subsection{Stellar content: point-like sources}\label{point_sources}
\begin{figure*}[!t]
  \includegraphics[width=\textwidth]{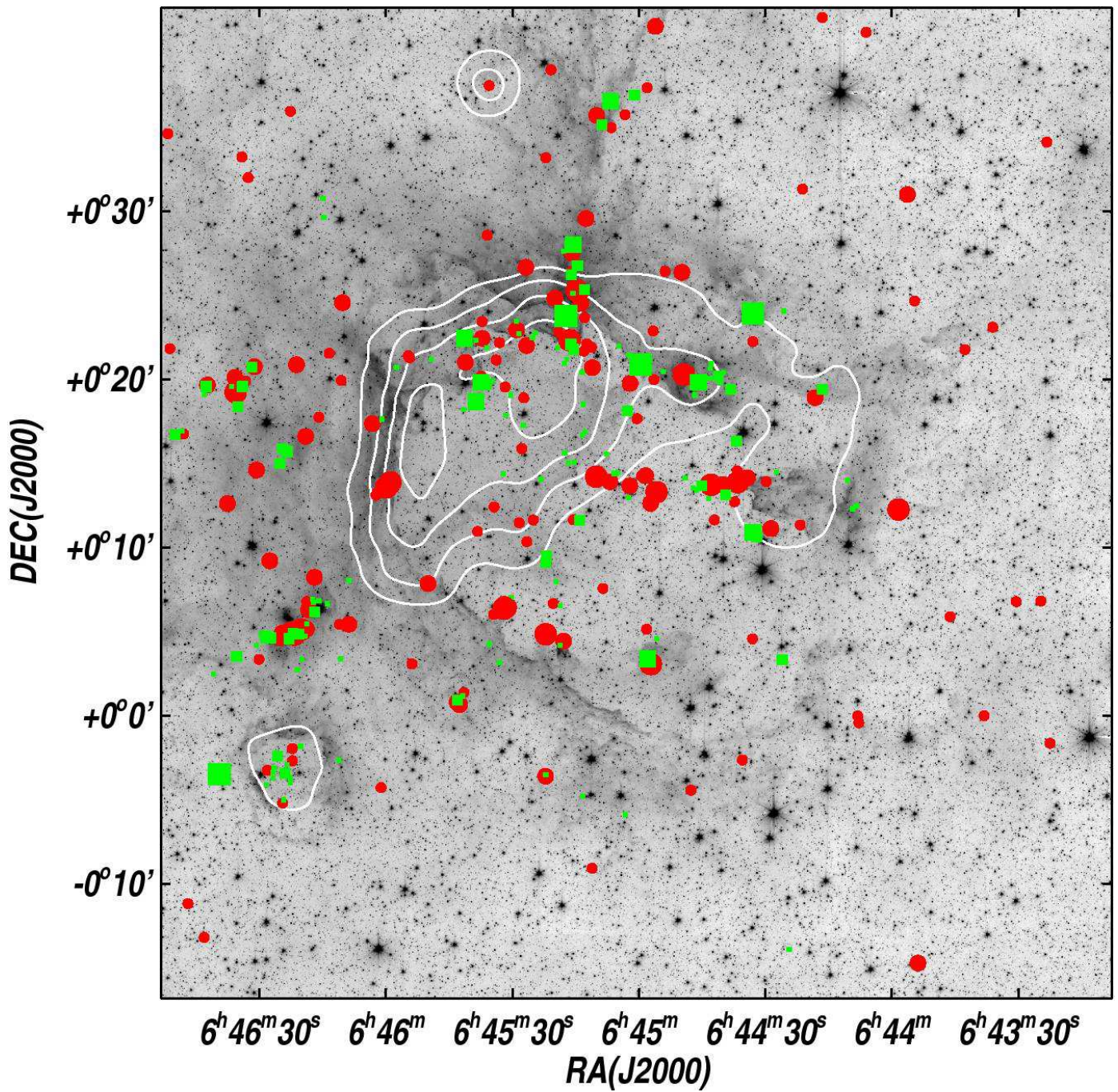}
  \caption{ Distribution of young stellar objects in the field around
    S284 on the IRAC\,3.6 map. Class~I objects are represented by
    filled red circles and scaled in size with the IRAC\,4.5
    brightness in the range 8.2--17 mag. Class~II objects are depicted
    with filled green squares and scaled in size with the IRAC\,5.8
    magnitude in the range 7.9--14.35 mag . Open circles correspond to
    possible Class~I objects that appear saturated in IRAC\,8 $\mu$m.
    White contours represent the radio continuum emission detected
    from the 4850MHz survey \citep{Condon_91}[see the electronic
    edition of the Journal for a colour version of this figure].}
  \label{yso_loc}
\end{figure*}
\begin{figure}[!t]
    \includegraphics[width=8.8cm]{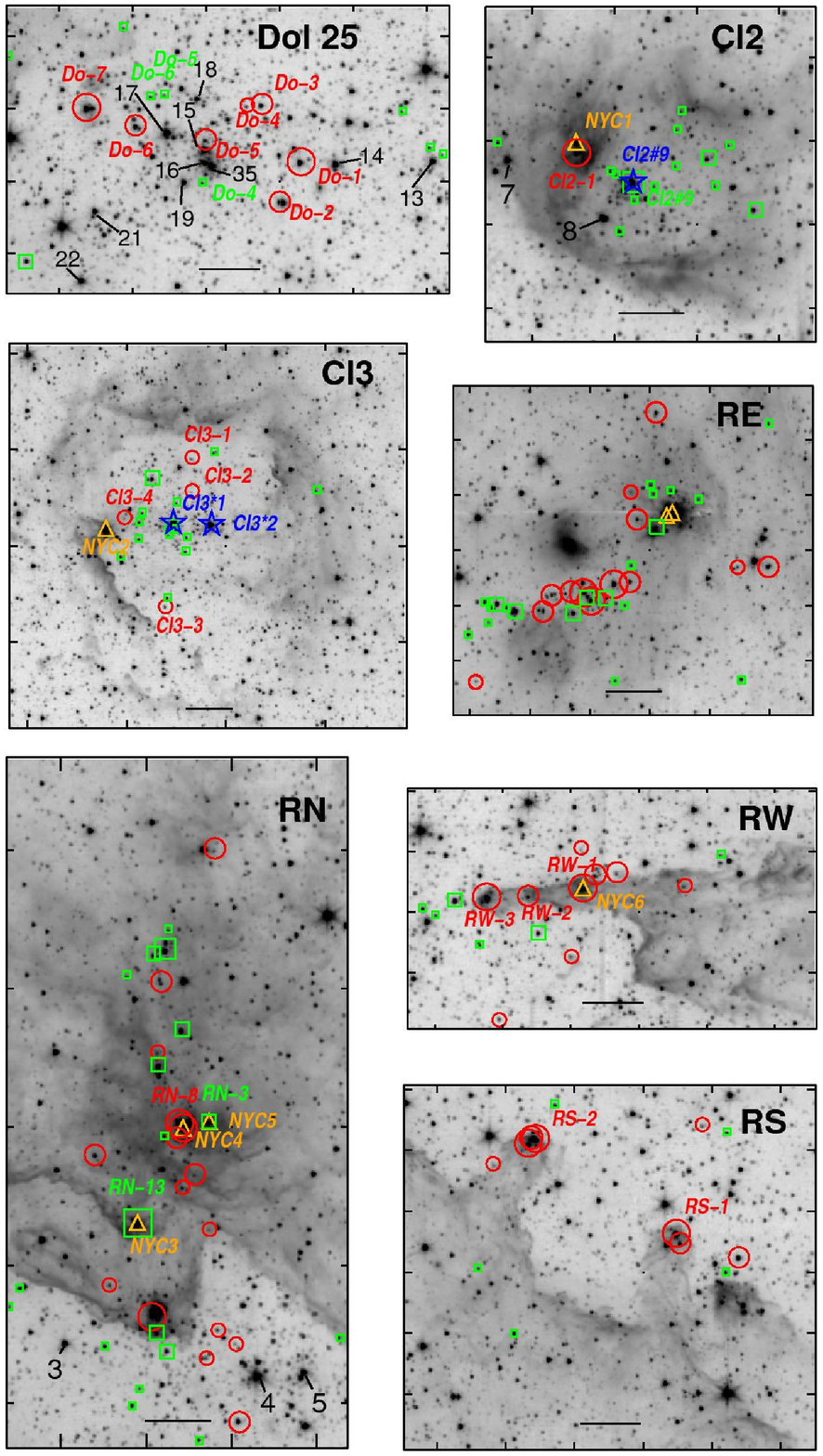}
    \caption{ Individual regions in S284 overlayed on the IRAC\,3.6
      map. Symbols correspond to Class~I objects ({\it red circles}),
      Class~II objects ({\it green squares}) and near-IR excess
      sources ({\it orange triangles}). Blue stars mark the location
      of the powering stars in Cl2 and Cl3. Horizontal bars represent
      a scale of 1$\arcmin$. Black labels correspond to the source ID
      given in \cite{Moffat_Vogt} [see the electronic edition of the
      Journal for a colour version of this figure].}
  \label{ir_mosaic}
\end{figure}
Our IRAC/Spitzer data has unveiled the presence of a rich stellar
population in S284 in the form of various peripheral clusters. It also
allows us to put constraints on the debated extent of the cluster
members of Dolidze 25 through the link between cluster and larger dust
structures. Another important point is the inferred presence of
high-mass stars in clusters Cl2 and Cl3 through the detection of
associated ionized regions, which requires the existence of at least
one star of spectral type earlier than a B3V \citep{Churchwell02}.

Our mid-IR IRAC data allows the classification of point-like sources
based on the differentiation between sources that show an IR excess
and those that are photosphere-dominated. Furthermore, it is also used
as a young stellar object subclassification between Class~I and
Class~II objects by comparison with models of proto-stellar envelopes
and circumstellar disks \citep{Allen04}. The effective temperature and
luminosities of young stars can be used to infer their ages and thus
the timescale for the different evolutionary stages of the
circumstellar material. Although the literature comprises information
on the spectral types of a small sub-sample of bright stars
\citep{Lennon,Moffat_Vogt,Babu}, the reported spectral types vary
significantly; for instance, the spectral type clasiffication of the
brightest source in Cl2 --an A0 star according to \cite{Babu}-- is not
sufficient to produce the observed \HII{} region. As an alternative, a
crude estimate of the spectral type can be derived from the stellar
brightness in the IRAC\,4.5.

\subsubsection{IRAC/Spitzer colour-colour diagram}\label{irac_diag}
Our IRAC observations open a new window on the stellar content in
S284. This improvement does not come so much from peering through the
extinction at longer wavelengths, but from the fact that the
sensitivities achieved in IRAC\,3.6 and IRAC\,4.5 are higher than
existing near-IR surveys. Our mid-IR colour-colour stellar
classification is restricted by the limiting magnitudes in IRAC\,5.8
and IRAC\,8.0 (14 and 13.5 mag, respectively), since simultaneous
detection in the four bands is needed for this classification method.
In addition, to increase the reliability of our classification, we
only consider those objects with photometric error smaller than 0.175
mag in each of the four bands.

The diagram using a total of 4046 sources is presented in
Fig.~\ref{CI-CII_total}. Bona fide candidate Class~I objects are
selected from the region marked in the diagram with the restriction
that [3.6]$-$[4.5]$>$0.4 mag. The bulk of main sequence and giant
stars is centred at a [3.6]$-$[4.5]$=$\,$+$0.1 mag, thus on average
slightly bluer than expected. As a consequence, main sequence and
giant stars contaminate the area of the diagram where Class~II objects
lie according to the models \citep{Allen04}. For this reason, we have
designated as Class~II sources those whose [3.6]$-$[4.5] colour is
above the visible gap in this direction, namely [3.6]$-$[4.5]$>$\,0.3
mag. A particular group of sources lies in a region of the diagram
where [5.8]$-$[8.0]$>$\,1.4 mag and [3.6]$-$[4.5]$<$\,0 mag. These
points correspond to enhancements in the filamentary extended
structure that MOPEX detected as point-like sources whose colours
correspond to reflected emission, i.e., reflection nebulae.

The full lists of Class~I and Class~II candidates are available
electronically in Tables~3 and 4.

\subsubsection{2MASS diagrams}
The complementary information provided by the 2MASS catalog is
especially useful for the brightest stars that appear saturated in our
IRAC images. This is the case for bright central sources located in
the heads of a few cometary globules. It is important, however, to
bear in mind the slightly lower spatial resolution of the 2MASS
survey, especially toward the centre of the more dense clusters. The
near-IR colour-magnitude and colour-colour diagrams of the sources
within S284 are displayed in Fig.~\ref{NIR_CMD_CCD}.

For the colour-magnitude diagram (CMD), we consider sources with
relative photometric error below 20\% in the three 2MASS bands. The
main sequence (MS) at a distance of 5 kpc with A$_{V}=0$ and A$_{V}=5$
mag are displayed for reference. Note that the estimated average
extinction toward Dolidze 25 corresponds to A$_{V}$ $=$ 1.98 mag
\citep{Turbide_Moffat}. The M$_\mathrm{V}$ magnitudes of luminosity
class V are obtained from \cite{Schmidt-Kaler}. For the colours of the
spectral types earlier than A0V we use the values from \cite{Wegner}
and for later spectral types we obtained the colours from
\cite{Tokunaga}. The near-IR extinction law adopted is from
\cite{Rieke&Lebofsky}. We select the sources redder than the gap
noticeable at $J$ $-$ $K_s$ $\sim$\,$-$0.3 mag in our CMD and brighter
than an accordingly reddened star B8V (points represented in red in
Fig.~\ref{NIR_CMD_CCD}). These objects are either intermediate- and
high-mass stars that are strongly reddened or more evolved stars that
have reached the giant phase. These presumably reddened stars are
plotted in the colour-colour diagram (CCD) shown in
Fig.~\ref{NIR_CMD_CCD} ({\it right panel}), where many of them lie in
the red giant locus.

In our CCD we consider that stars posses a near-IR excess when they
appear redward of the region limited by the reddening vectors
projected from the intrinsic colours of main sequence stars and giants
and above the T~Tauri locus \citep{Meyer}. These objects (marked with
triangles in the diagram) are considered to be YSO candidates
\citep{Ojha_04}. The same triangle-like symbols are used to highlight
those objects back on the CMD.

\subsubsection{The distribution of YSOs}\label{dist_sources}
The spatial distribution of YSO candidates in S284 is shown in
Fig.~\ref{yso_loc}. The symbol sizes for the Class~II candidates are
scaled with the IRAC\,4.5 brightness as a proxy for the luminosity of
the YSO \citep{Whitney_04}. This choice is motivated by the fact that
the PRF is undersampled in the IRAC\,3.6 band, while oversampled in
the other bands; IRAC\,4.5 thus offers a more reliable measure of the
flux. Moreover, the extended emission appears least prominent in the
4.5~$\mu$m map. Analogously, the symbol sizes for Class~I objects are
proportional to the IRAC\,5.8 brightness. Although IRAC\,8.0 seems
better suited to YSOs with a strong excess, the 7.7\,$\mu$m PAH
emission band could contribute significantly to the source flux
estimate \citep{Peeters}.

Most Class~I and Class~II candidates arise in clusters around the
shell of the B1 dust bubble or at the locations of the powering
clusters Dolidze~25, Cl2, and Cl3. Other YSOs around B1 seem to be
more scattered, but for some cases their low brigthness make them
dubious. Several bright Class~I objects are located at the head of
cometary globules or at elephant trunks around B1. In particular, the
longest trunk of the region harbours several Class~I candidates at
various positions along the trunk (RW in Fig.~\ref{box_blobs}). Those
candidates at the rim of the B1 bubble show a colour
[3.6]$-$[4.5]$>$\,1.1 mag indicative of either strong reddening or
high stellar temperatures \cite{Whitney_04}. We also detect two
clusters of Class~I and Class~II objects (RE in Fig.~\ref{box_blobs}
and another concentration 15$\arcmin$ north of RE) a few parsec
distant from the ionization front associated with B1. To the north of
the IRAC field, we find one isolated Class~I candidate at the centre
of a region of strong radio continuum emission.

Surprisingly, Dolidze~25 comprises several Class~I objects located in
a 4$\arcmin$ projected diameter region. This result will be discussed
further in Sect.~\ref{age+timesc}. Cl2 and Cl3 seem to have a small
population of YSOs dominated by Class~II objects, whereas a Class~I
candidate is detected at the rim of each bubble. In conclusion, the
location of the YSO candidates follows closely the morphology of the
nebulae. There is a higher concentration of young objects along the IR
bright rim of B1. However, a significant number of objects are
detected as far as few tens of parsecs away from the rim of B1.

\subsubsection{Comments on the stellar population of individual regions}
\begin{figure*}[!t]
  \includegraphics[angle=90,width=\textwidth]{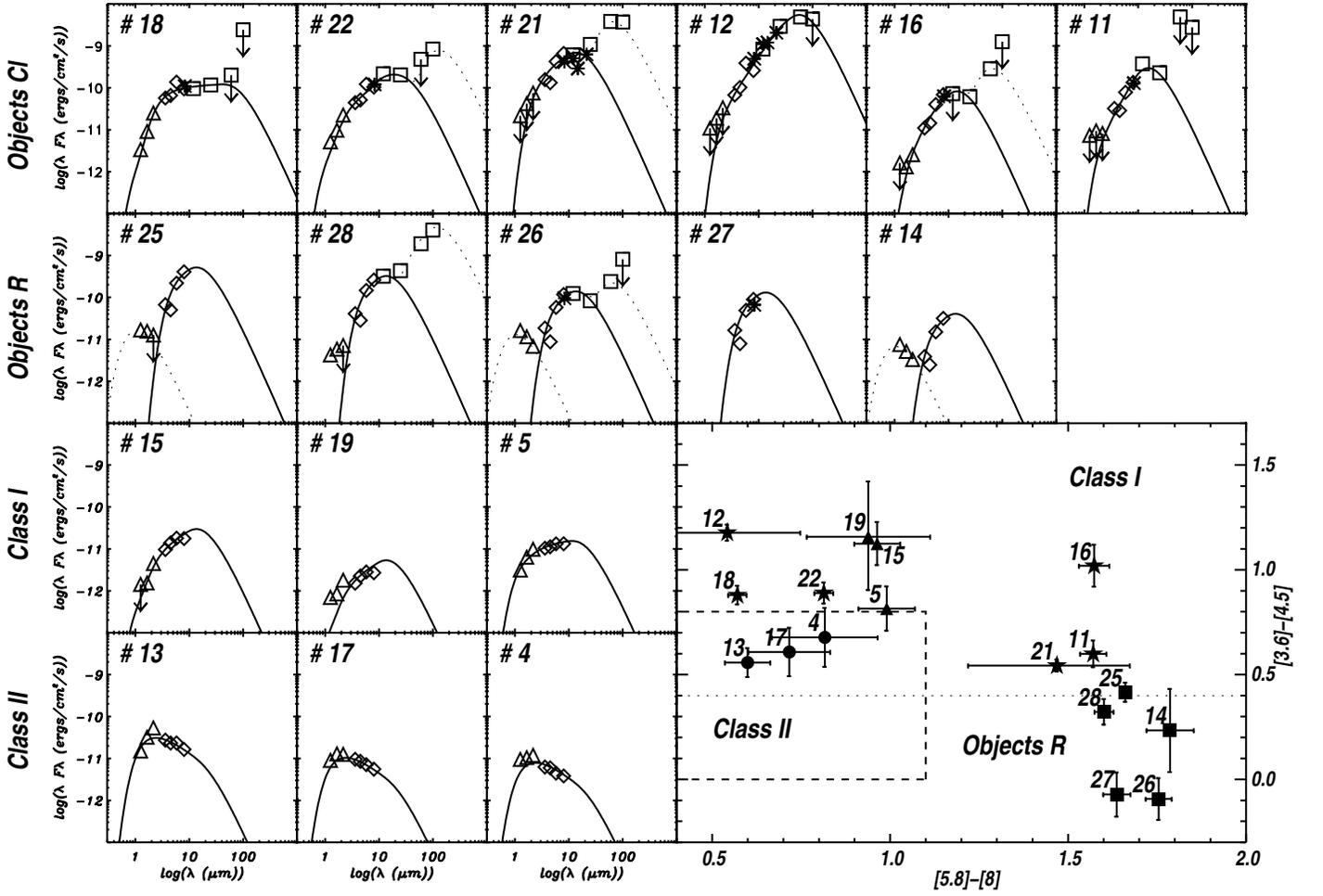}
  \caption{SEDs of the 11 blobs identified in the field of S284. The
    solid lines represent the total black-body distribution taken into
    account to compute the luminosity. The dotted lines show
    additional components required to better match the overall
    distribution. (Bottom-right panel) IRAC colour-colour diagram of
    the blobs. Symbols correspond to Class~I unresolved objects ({\it
      triangles}), Class~II unresolved objects ({\it circles}), blobs
    with reflection colours ({\it squares}), and blobs with colours of
    an early cluster phase ({\it stars}).}
\label{seds}
\end{figure*}
\begin{figure*}[!t]
  \includegraphics[width=\textwidth]{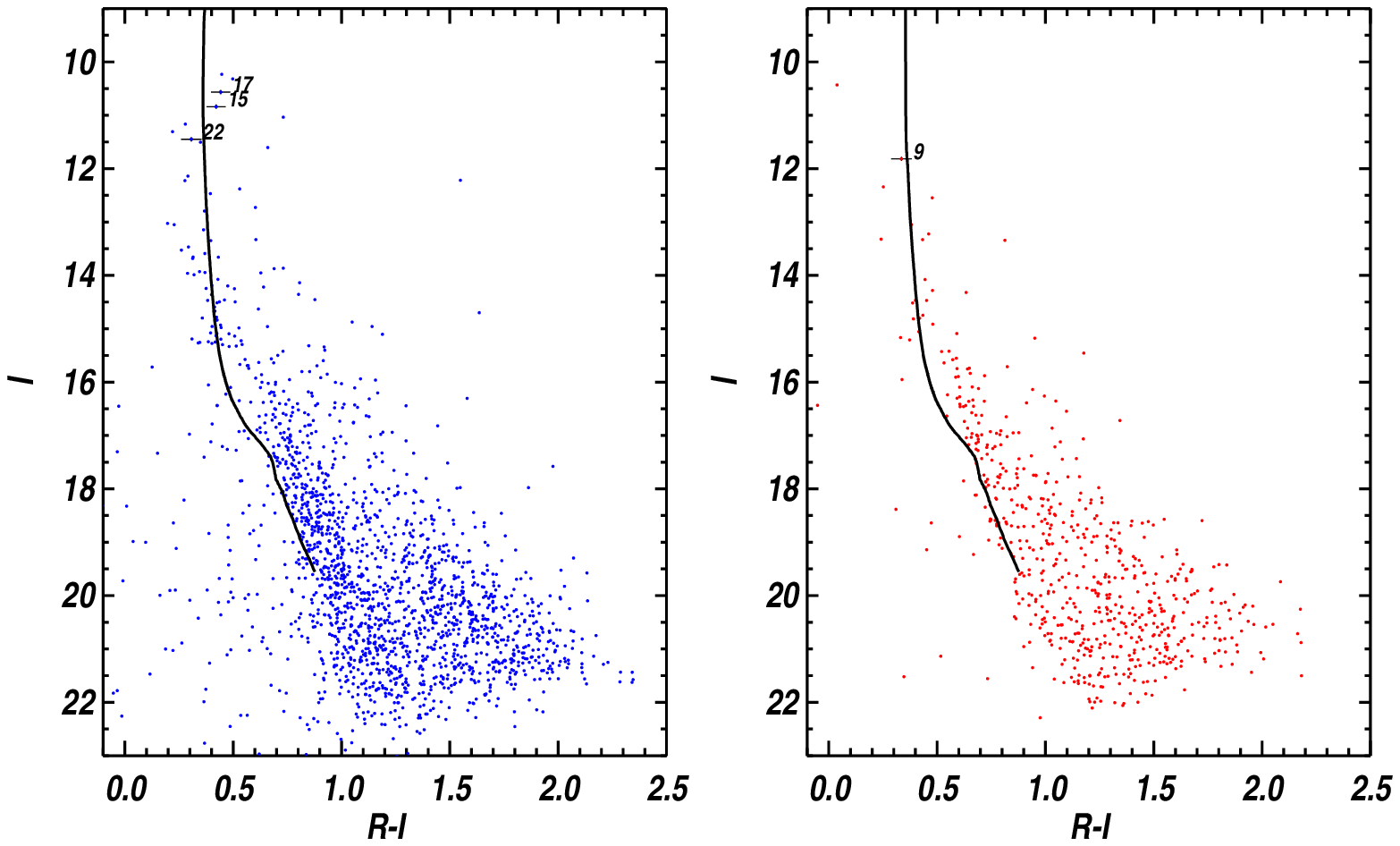}
  \caption{{\bf (Left panel)} Colour-magnitude diagram of the field
    around Dolidze 25 combining VIMOS and INT/WFC photometry
    \citep{Cusano,deleuil}. The overplotted isochrone corresponds to a
    log(age[yr])$=$6.75 and Z$=$0.004 \citep{Lejeune_01}. The distance
    modulus corresponds to 14.4, while E($R-I$)$=$0.5 mag. {\bf (Right
      panel)} Analogous colour-magnitude diagram for Cl2 with an
    overlayed isochrone of log(age[yr])$=$6.0. [See the electronic
    edition of the Journal for a colour version of this figure].}
\label{optical_cmd_iso}
\end{figure*}
\paragraph{Dolidze~25}
Most of the objects that have been traditionally considered as members
of Dolidze~25 (e.g. \#17, \#15, \#16, \#35, \#22, \#18, \#19) do not
show any circumstellar peculiarity at near- and mid-IR wavelengths
(see Fig.~\ref{ir_mosaic}). A concentration of Class~I candidates
exists toward Dolidze~25, while low-luminosity Class~II sources appear
more scattered around the cluster. Most of the Class~I candidates are
also detected in the $J$ band ruling out the possibility of being
reddened Class~II objects.

\paragraph{Cl2}
The area around Cl2, also including B2, shows a high concentration of
Class~II objects. The source located at the centre of the cluster
(source \#9 in Table~\ref{table_st}) has been spectroscopically
classified in the optical as an O9V star \citep{Moffat_ST}. The
cluster core appears elongated at mid-IR wavelengths and starts to be
resolved redward of 5.8 $\mu$m. The counterpart to O9V star exhibits
the mid-IR colours of a Class~II object, although this could be due to
the contamination by the nearby object. Another source observed in the
direction of the \HII{} region (\#8) appears extended in our IRAC\,3.6
image and is barely resolved into two point-like sources in the redder
IRAC maps. A bright Class~I object is detected at the rim of the dust
bubble, where the extended emission is the strongest. Another nearby
object shows a strong near-IR excess (NYC1).
 
\paragraph{Cl3}
The two brightest objects within the central cluster (Cl3*1 and Cl3*2)
are detected in all four IRAC bands, but their SEDs only show a
photospheric component. Their location in the near-IR CMD suggests
that these are O-type stars. In particular, Cl3*1 is closely
surrounded by sources that show a mid-IR excess. NYC-2 is located at
the head of the cometary globule east of bubble B3 and appears
saturated in IRAC\,3.6, IRAC\,4.5, and IRAC\,8.0, so it is not
included in our analysis of Sect.~\ref{irac_diag}. However, the
overall shape of its SED corresponds to a Class~I candidate.

\paragraph{RN}
The field labelled as RN includes several cometary globules to the
north of B1. Two of the brightest sources at the vortices of cometary
globules appear saturated in one of our IRAC maps, and yet, bright
Class~I objects are unveiled within these clusters, pointing to the
presence of various YSOs embedded in these regions. The detection of
two Class~II candidates (RN-13 and RN-3) is reinforced by the near-IR
excess shown in the 2MASS photometry (sources NYC3 and NYC5). One of
them coincides with the head of a non-saturated globule that hosts
only few point-like objects.
 
\paragraph{RE}
The field RE has two blobs of strong emission separated by a projected
distance of $\sim$0.4 pc. The eastern blob is resolved into extended
emission, while the western blob has several resolved point-like
sources, the central one saturated. The 2MASS photometry toward this
source reveals a near-IR excess at a slightly worse resolution than
our IRAC images. A resolved cluster is detected to the south of the
blobs. Seven bright Class~I candidates are aligned in a trail, while
several Class~II sources are scattered around. This cluster is
partially detected in the near-IR and optical, indicating that it is
not deeply embedded within the dust; hence, we can rule out that the
Class~I candidates are reddened Class~II objects. Moreover, these
Class~I candidates are the only objects not located at the rim of B1
that show a colour [3.6]$-$[4.5]$>$\,1.1 mag, pointing to relatively
high luminosities \citep{Whitney_04}, but not hot enough to generate
an \HII{} region.

\paragraph{RW and RS}
The region RW encompasses the longest trunk at the rim of B1. Several
Class~I objects are located along the pillar. One of them is also
detected by an associated near-IR excess.

Analogously, the RS region contains another two trunks that harbour
Class~I sources at the head of their cometary globules.

\subsection{Stellar content: embedded objects and blobs}\label{blobs}
The previous sections have focused on the stellar classification of
point-like sources in S284. However, the filamentary high-density
structures detected in our IRAC maps support the possible presence of
embedded clusters in regions where the emission appears resolved.

We selected 11 blobs or regions of small extent (a sub-scale of
bubbles B2 and B3) and high infrared brightness in order to search for
young embedded clusters. The radius of these blobs was defined as 1.25
$\times$ HWHM in the IRAC\,8.0 band and kept constant to calculate the
integrated flux in other bands. The same radius was used to search for
near-infrared counterparts in the 2MASS point-source database. When
multiple sources were present, we summed their contribution to
estimate the total near-infrared flux. For the objects that saturated
an IRAC-band image, we considered the IRAC fluxes as a lower limit and
MSX fluxes when available. We also searched for far-infrared
counterparts in the IRAS point sources catalog and selected these that
are compatible with the blobs as their main contibutors in the IRAS
beam on the basis of HIRES images. As expected, the more extended
blobs are more likely to have an IRAS counterpart.

All these data points were used to build the spectral energy
distributions of the blobs presented in Fig.~\ref{seds}. We fitted the
2MASS, IRAC, MSX and IRAS flux distribution with a combination of
blackbodies with temperature ranging from 30 to 1000\,K to reproduce
the shape of the SEDs and estimate their corresponding luminosities.
Figure~\ref{seds} displays the spectral energy distribution as well as
the results of the fit. The blob characteristics are described in
Table~\ref{blob_table}, while their location and radius are
represented in Fig.~\ref{box_blobs}.

We identify four different types of SEDs in column (6) of
Table~\ref{blob_table}: (I) ClassI objects unresolved in IRAC\,8.0 and
with integrated near-infrared luminosities $<$ 2.0; (II) ClassII
objects, also all unresolved in IRAC\,8.0; (R) objects having
reflection colours; and (Cl) objects falling in the Class I region in
the IRAC colour-colour diagram with higher near-infrared luminosities.
This last category also tends to have a steeper 3.6 to 8 $\mu$m slope
than unresolved ClassI objects, nearly as steep as reflection objects
for some of them. Still, their mid-infrared spectrum cannot be fitted
with a single temperature blackbody, but on the contrary require an
extended range of temperatures. Their luminosity, extent, and spectral
distribution suggest that these are young clusters, either
constituting an earlier cluster phase compared to bubbles B2 and B3 or
lower-mass clusters.

\section{Discussion}\label{disc}
\subsection{Environmental conditions}\label{environ}
Before we consider the stellar content and the ages of the various
distinct regions, it is instructive to debate the global morphology
and what can be concluded from it. The shell thickness of dynamically
blown bubbles is a function of the central star, the age of the
bubble, and the density of the ambient ISM \citep{Weaver_77}. The
theory predicts that the ratio of shell thickness to radius increases
with age and ambient interstellar density. Therefore, for relatively
symmetric bubbles with the powering star(s) located at the geometrical
centre (as is the case of S284), a smaller ratio is indicative of a
lower ambient density. We observe that the ratio of shell thickness to
radius is not constant around B1, decreasing along the southwest
direction. A lower ambient density in this direction would explain
this effect and support our interpretation of the lower dust
extinction values inferred from the IRAS maps as caused by a local
ambient density gradient in this direction.

The presence of elephant trunks suggests that the dynamics of the
\HII{} region within B1 are dominated by the thermal pressure of the
gas at 10$^4$ K and not by shocked gas, either produced by stellar
winds or supernovae. These sawtooth shapes have been successfully
modelled as the result of dynamic instabilities in the ionization
front of an expanding \HII{} region, regardless of pre-existing
density inhomogeneities, by \cite{Garcia-Segura}. These models account
for different cooling and background conditions and outline the effect
of low-metallicities (final shell temperatures are higher) into
thicker shells and wider and sparser trunks.

Some characteristics of this SFR may be determined by the combined
effect of the lower than solar metallicity and a gas pressure that is
lower than typically found in the inner MilkyWay. High-mass stars
exhibit weaker winds at lower metallicity due to the less efficient
momentum transfer of the stellar radiation on the outer envelope. In a
study of young SFRs in the SMC, \cite{Sabbi_08} find that the
morphology of the region is not shaped by wind interaction but more by
the expansion of the ionized bubble, much like what is found for S284.
A second effect concerns the dust shielding. Low-metallicity
environments like the LMC and SMC show a deficit of the smallest dust
grains \citep[e.g.][]{Stanimirovic}, which may cause the UV radiation
of the newly formed stars to permeate the surrounding molecular cloud
and modify the conditions over great distances (e.g. dust composition,
charge state of the smallest grains, photodissociation of molecular
gas, Hony et al. in preparation). It is interesting to compare this
with the results of \cite{Boissier_07}. These authors study the star
formation in a sample of late-type galaxies using IR and UV and find
that the star formation in the outskirts of these galaxies is typified
by less attenuation due to the lower metallicity and that these
regions are underluminous in H$\alpha$. One possible explanation for
the lack of prominent H$\alpha$ emission is that the lower ambient
pressure makes the average star-forming cluster less massive, and as a
result less very massive stars are formed. S284 might be an example of
such a less massive star-formation cluster, with an O7V the most
massive star detected.

\subsection{Spectral types and masses}\label{spect+mass}
The lack of detailed information on the spectral types of the complete
stellar content precludes a thorough age analysis of S284. The use of
spectral types has proven useful for determining age gradients in
massive clusters like the Eagle nebula \citep{Martayan} and probing
the disk evolution toward several Galactic SF regions
\citep[e.g.][]{Sicilia-Aguilar}. However, we qualitatively discuss
this issue with the information presently at hand.

Table~\ref{table_st} compiles the spectral type and luminosity class
classification and membership for S284 from the literature. The
various maps tracing the dust and ionized gas complement this existing
information on S284 distinguishing among the different physical
structures (clusters, \HII{} regions and bubbles).

\subsection{Relative ages and timescales}\label{age+timesc}
\begin{table*}[!t]
    \caption{Characteristics of the selected embedded regions and
      comparison with Class I and II unresolved objects.} 
    \small
    \begin{tabular}{lccccccc}
      \hline\hline
      Blob ID & RA & DEC & Radius&Log(L/Lo)&Class&Region&IRAS assoc.\\
      &    &     &  ('') &         &     &      &           \\
      \hline							   
      18..............    & 06  45  40.8 & +00  22  27.1 &  5.73 &2.5 &Cl  &  &06430+0025\\ 
      22..............    & 06  46  29.6 & -00  03  29.2 &  6.06 &2.6 &Cl  &Cl3&06439-0000\\
      21..............    & 06  46  15.8 & +00  06  26.2 & 15.92 &3.0 &Cl  &RE&06437+0009\\
      12..............    & 06  45  16.0 & +00  22  26.9 &  8.24 &3.9 &Cl  &RN&06426+0025\\
      16..............    & 06  45  31.8 & +00  06  26.1 &  5.31 &2.0 &Cl  &RS&06430+0009\\
      11..............    & 06  45  14.1 & +00  25  18.6 & 10.86 &2.6 &Cl  &RN&06426+0028\\
      25..............    & 06  46  23.3 & +00  05  48.6 & 22.39 &2.9 &R   &RE&\\
      28..............    & 06  44  49.1 & +00  20  15.3 & 20.14 &2.7 &R   &Cl2&06422+0023\\ 
      26..............    & 06  44  31.8 & +00  23  23.6 & 13.22 &2.3 &R   &  &06419+0026\\ 
      27..............    & 06  45  18.9 & +00  03  49.2 & 12.80 &2.3 &R   &RS&\\ 
      14..............    & 06  45  17.5 & +00  03  30.3 &  5.04 &1.8 &R   &RS&\\ 
      15..............    & 06  45  22.0 & +00  04  49.4 &  U    &1.7 &I   &RS&\\ 
      19..............    & 06  45  58.8 & +00  13  52.9 &  U    &0.9 &I   &  &\\ 
      5................    & 06  44  36.5 & +00  13  51.2 &  U    &1.5 &I   &RW&\\ 
      13..............    & 06  45  17.3 & +00  23  45.6 &  U    &1.8 &II  &RN&\\ 
      17..............    & 06  45  37.4 & +00  19  49.8 &  U    &1.3 &II  &  &06431+0023\\ 
      4................    & 06  44  32.9 & +00  10  52.4 &  U    &1.2 &II  &RW&\\ 
      \hline
      \label{blob_table}
    \end{tabular}
\end{table*}
The different spatial scales of \HII{} regions and their distribution
around S284 is suggestive of different epochs of star formation.
Nevertheless, a sequence of events and causality between them must be
demonstrated to confirm a triggered star-formation scenario.
 
Low-mass Class~I objects have an estimated age of a few 10$^5$ yr,
while Class~II objects range between 10$^6$ and 10$^7$ yr
\citep{White}. We can therefore use these two classes of protostars
present in the S284 field as a relative age scale to study the
interaction of young clusters ($<$\,5 Myr) with their environment.

The three dust bubbles that bound the \HII{} regions detected on our
H$\alpha$ map (B1, B2, and B3) share a common morphology of cometary
globules or elephant trunks. In the case of B1 and B3, they are
clearly protruding into the ionized gas. The brighter young YSOs
(Class~I) and proto-clusters appear located at the vortices of these
structures, pointing to a later formation than that of the central
clusters in the \HII{} regions. That no compact radio continuum
emission, maser emission or strong CS have been detected at these
locations supports the idea that these young protostars and
proto-clusters are probably not massive. The morphology of this
younger generation of protostars and proto-clusters is compatible with
a process of dynamical instabilities of the ionization front
\citep{Garcia-Segura}. Because the condensations do not seem to share
the expansion of the ionized gas, another plausible explanation would
be the radiation driven implosion of pre-existing molecular
condensation. The morphology of the region at the rim of B2 where
several YSO candidates are found resembles the cases where the collect
and collapse process has be reported to be at work
\citep{Zavagno_06,Zavagno_07}. The velocity distributions of the
ionized and molecular gas have been proven valuable when studying the
association of condensations and YSOs with \HII{} regions
\citep{Pomares_09}.

B1 shows a thick dust shell with a combination of Class~I and older
Class~II candidates, both in clusters and in isolation, farther away
from the ionization front. \cite{Zavagno_07} report a similar result
in their study of the RCW 120 region, explaining it in terms of a leak
of radiation into the PDR.

A question arises after discussion of the impact of the various
clusters on their immediate surroundings: what is the interplay
between the  high-mass stellar populations in this region? are Cl3,
and especially Cl2, produced by the expansion of the \HII{} region
powered by Dolidze~25?

The powering clusters Cl2 and Cl3 are dominated by Class~II
candidates. It is therefore puzzling to detect younger Class~I
candidates in the direction of the central cluster Dolidze~25, the
presumed oldest cluster in this region. In the case that these YSO
candidates were associated to the cluster, these results would
challenge the triggered SF scenario initially suggested by the
multiple bubble morphology of S284, in particular the close location
of Cl2 to the B1 shell. The UV field produced by the several O-type
stars of Dolidze~25 must contribute to the photo-evaporation of
circumstellar material in nearby stars. However, YSO candidates have
also been detected close to the young massive cluster NGC~6611 in the
Eagle nebula \citep{Indebetouw_07}. The authors conclude that
circumstellar disks must be fairly robust to the cluster environment
around NGC~6611. Another possible explanation would be that these
Class~I objects are located at the border of the bubble, either in the
foreground or in the background. Further support for this hypothesis
could come from the detection of several Class~II candidates in
Fig.~\ref{yso_loc}. These Class~II sources would trace the stellar
population associated with a large emission filament that crosses B1
in the northeast direction in our IRAC images and appears in
absorption in our H$\alpha$ image. The proximity {\it in projection}
of this filament suggests that this or other structures in the line of
sight could overlap with the cluster Dolidze~25. If the Class~I
objects detected near Dolidze~25 were not members, the cluster would
be dominated by a stellar population of naked photospheres. A future
radial velocity study of the stars around Dolidze~25 would aid in
answering this question.

\subsubsection{Dynamical ages of the \HII{} regions}
\begin{table}[!t]
  \caption{Membership and Spectral Types of Dolidze~25. Column (1) indicates the star number following the notation of \cite{Moffat_Vogt}. (m) members, (n) non-members.}
  \label{table_st}
  \renewcommand{\footnoterule}{}
  \small
  \begin{tabular}{l@{\ }c@{\ }c@{\ }c@{\ }c@{\ }c@{\ }c@{\ }c}
    \hline\hline
    ID   &  \multicolumn{7}{c}{Classification}                                          \\
    (1) & (2) &  (3)  &   (4)   & (5)&     (6)     &       (7)          &  (8)         \\
    \hline
    1   & (m) & B0IV  & B5(m?)  &    & B5          &                     & (n)         \\
    2   & (n) &       & A5(--)  &    & A5          &                     & (n)         \\
    3   & (n) &       & A0(--)  &    & A0          &                     & (n)         \\
    4   & (n) &       & B2(--)  &    & F5$^{\dag}$ &                     & (n)         \\
    5   & (n) &       & B5(--)  &    & A0$^{\dag}$ &                     & (n)         \\
    6   & (n) &       & A5(--)  &    & A5          &                     & (n)         \\
    7   & (n) &       & A0(--)  &    & A0$^{\dag}$ &                     & (n)         \\
    8   & (n) &       & A5(--)  &    & A2$^{\dag}$ &                     & (Cl~2 m)    \\
    9   & (m) & O9    & A0(m)   &    & A0          &                     & (Cl~2 m)    \\
    10  & (m) & F2Iab & F0(m)   &    & F5$^{\dag}$ &                     & (n)         \\
    11  & (n) &       & K2      &    & K2$^{\dag}$ &                     & (n)         \\
    12  & (m) & B0III & B2(m)   &    & B2          &                     & (Dol~25 m?) \\
    13  & (m) &       & B8(m)   &    & B8          &                     & (Dol~25 m?) \\
    14  & (n) &       & B8(--)  &    & B8          &                     & (Dol~25 m)  \\
    15  & (m) & B0III & B8(m)   & (m)& B8          &O6$^{\ddag}$         & (Dol~25 m)  \\
    16a & (x) &       &         &    &             &                     & (Dol~25 m)  \\
    16b & (m) &       &         & (m)& B8          &                     & (Dol~25 m)  \\
    17  & (m) & B1V   & B8(m)   & (m)& B8          &O5.5--O6.5$^{\ddag}$ & (Dol~25 m)  \\
    18  & (m) &       & F5(m)   & (m)& F5          &                     & (Dol~25 m)  \\
    19  & (m) &       & A5(m)   &    & A5          &                     & (Dol~25 m)  \\
    20  & (n) &       & F5(--)  &    & F5          &                     & (Dol~25 m)  \\ 
    21  & (n) &       & F5(--)  &    & F5          &                     & (Dol~25 m)  \\
    22  & (m) &       & A5(m)   & (m)& A5          & O8--O9$^{\ddag}$    & (Dol~25 m)  \\
    23  & (n) &       &         &    &             &                     & (n)         \\
    24  &     &       & B8(m?)  &    &             &                     & (n)         \\
    25  &     &       &         &    &             &                     & (n)         \\
    26  &     &       & K0(--)  &    &             &                     & (n)         \\
    27  &     &       & F5(m)   &    &             &                     & (n)         \\
    28  &     &       & F5(m)   &    &             &                     & (n)         \\
    31  &     &       & F0(m)   &    &             &                     & (n)         \\
    32  &     &       & F0(m)   &    &             &                     & (n)         \\
    33  &     &       & B8(m)   &    &             &                     & (n)         \\
    34  &     &       & F5(m)   &    &             &                     & (n)         \\
\hline
\end{tabular}
\\
(2) \cite{Moffat_Vogt},
(3) \cite{Moffat_ST},
(4) \cite{Babu},
(5) \cite{Turbide_Moffat},
(6) WEBDA Open Cluster Database http://obswww.unige.ch/webda,
(7) \cite{Lennon}
(8) This work,
$^{\dag}$ Spectral type classification from \cite{Cannon},
$^{\ddag}$ Spectral type determination following \cite{Martins_05}
\end{table}
The exceptional symmetry of S284 allows a rough calculation of the
dynamical age of the \HII{} region using a zero-order approximation
(i.e. a simple pressure driven expansion), using the strong shock
approximation and assuming a constant-density medium. We convert the
spectral types of the most certain members from the literature (see
Table~\ref{table_st}) into the corresponding Lyman continuum photon
rate. Dolidze~25 and its associated \HII{} region are considered
separately from Cl2 and its compact ionized region. Adopting an
isothermal sound speed $\sim$0.4 km\,s$^{-1}$, a recombination
coefficient $\sim$3 $\times$10$^{-13}$ cm$^3$\,s$^{-1}$ and the
estimated electronic density n$_e$ $\sim$ 100 cm$^{-3}$
\citep{Shaver}, we derive a dynamical age $<$\,3.8 Myr for B1. In the
case of the compact \HII{} region bound by B2, we assume n$_e$ $\sim$
2$\times$10$^3$ cm$^{-3}$ and a sound speed $\sim$0.4 km\,s$^{-1}$,
rendering a dynamical age $<$\,1.8 Myr.

\subsubsection{Comparison to isochrones and evolutionary tracks}
We attempted a quantitative age analysis of the clusters Dolidze~25
and Cl2. For this purpose, we converted the photometry of the bright
stellar population obtained with INT/WFC to the standard
Johnson-Cousins photometric system (Deeg et al., private
communication). We combined these data with the VIMOS photometry in
the $R$ and $I$ bands of the faint population \citep{Cusano}. The CMDs
of both regions are presented in Fig.~\ref{optical_cmd_iso}. The
population around Dolidze 25 exhibits a main sequence, and even a hint
of lower-mass pre-main sequence in the CMD. This is not the case for
Cl2, where no distinctive features are detected in the CMD probably
due to the poor statistics.
The field around Dolidze 25 is not fully covered to the north of the
cluster by the VIMOS observations. This could explain the void of
sources detected at $I\sim$16 mag in the left panel of
Fig.~\ref{optical_cmd_iso}.  The fit of a log(age[yr])$=$\,6.75
isochrone to the blue part of the main sequence renders an
E($R-I$)$=$\,0.5 mag, considering a distance of 5 kpc.

The  atmospheric parameter determination for  source \#22
\citep{Lennon} secures an age upper limit for the cluster of
6.2$\pm$0.6 Myr (assuming a coeval stellar population). This estimate
was obtained  via comparison  to a 15 M$_{\sun}$-evolutionary track
with Z=0.004 from \cite{Lejeune_01}, although source \#22 is located
at the outskirts of Dolidze 25, while sources \#15 and \#17 belong to
the high-stellar-density core of the cluster. Source \#15 is
spectrally classified as a middle O-type dwarf star, while star \#17
appears as a more evolved hot star of spectral type III or Ib
\citep{Lennon}. Their positions in the CMD reveal the presence of
differential extinction toward the centre of Dolidze 25, detected also
in the $V$ vs. ($B-V$) diagram. Since the masses derived from
atmospheric models are particularly uncertain for stars hotter than
35000 K, no comparison to an evolutionary track is attempted.
Nevertheless, the  presence of an O6V star  (\#15) confirms that
Dolidze 25 is younger than 5 Myr.

The optimal isochrone fit  to the stellar population of Cl2 in
Fig.~\ref{optical_cmd_iso}{\it (right panel)} indicates that it is
also approximately 5 kpc away.  We  use the less certain spectral type
classification of source \#9 as  O9V \citep{Moffat_ST} and the
comparison to a 15 M$_{\sun}$-evolutionary track to establish that the
cluster must be younger than 1.0$\pm$1.0 Myr.

These results for Dolidze 25 and Cl2 are in good agreement with the
dynamical ages estimated in the previous section from  the
correspondent \HII{} regions.  Moreover, in the case of Cl2, the
detection of several Class~II objects within the cluster is compatible
with the estimated age.

The age estimates for the clusters Dolidze~25 and Cl2 are consistent
with the inferred dynamical ages of their respective \HII{} regions.
This translates into an age spread of approximately 3 Myr between the
two clusters. Considering the $\sim$20 pc separation between them,
S284 lies on the age spread--linear size relationship described by
\cite{Elmegreen_00} for the LMC clusters. This correlation is
explained as the result of the rapid formation of self gravitating
sub-clumps of a larger cloud complex. In this context, these results
hint that the formation of the massive clusters in S284 is not
necessarily triggered.

\section{Conclusions}\label{concl}
In this paper we have presented newly obtained IRAC/Spitzer and
H$\alpha$/INT observations of S284. These observations show --at a
very detailed level-- a relatively undisturbed isolated site of
high-mass star formation. We presented an overview of the young
stellar content of this region and sketched a scenario for the
evolution of this region, considering its morphology.

Our IRAC maps show the newly discovered dust diffuse component of the
nebula, which allows the investigation of a multiple-bubble
morphology. Since this sort of structure has recently been explained
in terms of triggered star formation, we analyse the population and
morphologycal features as proxies for their relative ages.

The dominant photospheric component traced by the filters IRAC\,3.6
and IRAC\,4.5 allows the detection of a cluster at the centre of the
three \HII{} regions shown by our H$\alpha$ map. These clusters must
therefore harbour at least one high-mass star to blow such detectable
ionized regions. Considering the remarkable symmetry of the bubbles
and cluster centre positions, the morphologigal variations within the
field are discussed in terms of the environmental conditions. Density
variations along the field may play a major role in explaining the
observed features.

Based on an IRAC colour-colour criterium on the photometry of
point-like objects toward S284, we found a total of 155 Class~I and
183 Class~II candidates. Most of these young objects concentrate at
the shells of the bubbles. The extrapolation of this analysis to the
SEDs of integrated regions of extended emission yields the conclusion
that six more embedded clusters are also located in the dust shell.

We have detected relative younger Class~I candidates close to
Dolidze~25 than the dominant Class~II populations toward clusters Cl2
and Cl3. We considered a possible alignment of foreground or
background, young YSOs in the line of sight of Dolidze~25, as well as
a robust nature for low-mass disks, previously invoked for the case of
NGC~6611 \citep{Indebetouw_07}. We studied the dynamical ages inferred
from the \HII{} regions. These are in good agreement with the ages
inferred from the older stellar population of Dolidze 25 and Cl2
through comparison of optical photometry with isochrones and evolution
tracks. Based on the age-spread and linear-separation argument, we
discard the expansion of the B1 bubble as triggering the formation of
Cl2 and most probably of Cl3.

Reddened younger luminous objects (Class~I) are only detected at the
vortices of cometary globules or elephant trunks, while a mixed
population of Class~I and Class~II objects is present both in a
clustered and a more scattered mode farther away from the ionization
front inside B1. This suggested age gradient is even more pronounced
considering the dominant Class~II populations within Cl2 and Cl3 and
the presence of Class~I candidates at their respective shells. As a
result, we see indications of triggered star formation between the
powering clusters of the \HII{} regions and the physical structures at
their respective shells.

Processes such as radiation-driven implosion of pre-existing dense
clumps, dynamical instabilities of the ionization front, and the
collapse of the collected layer may explain the variety of
morphologies observed in this region. We conclude that the S284
fulfils the necessary conditions for an in-depth study of triggered
star-formation by expansion of its individual \HII{} regions.

\begin{acknowledgements}
  We thank M. Deleuil and the exoplanet team of CoRoT for providing
  their photometric data. We also thank the Spitzer helpdesk and F.
  Marleau for the discussions about IRAC photometry as well as J.
  Matute for endless discussions on the MOPEX pipeline. E. Puga kindly
  acknowledges the support of the GEPI, Observatoire de Paris during
  her stay. This publication makes use of data products from the Two
  Micron All Sky Survey, which is a joint project of the University of
  Massachusetts and the Infrared Processing and Analysis
  Center/California Institute of Technology, funded by the National
  Aeronautics and Space Administration and the National Science
  Foundation. This research has made use of the NASA/ IPAC Infrared
  Science Archive, which is operated by the Jet Propulsion Laboratory,
  California Institute of Technology, under contract with the National
  Aeronautics and Space Administration.
\end{acknowledgements}

\bibliographystyle{aa}
\bibliography{S284_Puga_bib.bib}
\end{document}